\documentclass[]{aastex631}

\usepackage{longtable}

\shorttitle{Exploring the small-scale magnetic fields of the solar analog KIC 8006161 using asteroseismology}
\begin{document}

\title{ Exploring the small-scale magnetic fields of the solar analog KIC 8006161 using asteroseismology}

\author{Guifang Lin}
\affiliation{Yunnan Observatories, Chinese Academy of Sciences, Kunming 650216, People's Republic of China}
\affiliation{International Center of Supernovae at the Yunnan Key Laboratory, Kunming 650216, People's Republic of China}

\author{Yan Li}
\affiliation{Yunnan Observatories, Chinese Academy of Sciences, Kunming 650216, People's Republic of China}

\affiliation{International Center of Supernovae at the Yunnan Key Laboratory, Kunming 650216, People's Republic of China}
\affiliation{University of Chinese Academy of Sciences, Beijing 100049, People's Republic of China}
\affiliation{Center for Astronomical Mega-Science, Chinese Academy of Sciences, Beijing, 100012, People's Republic of China}

\author{Jie Su}
\affiliation{Yunnan Observatories, Chinese Academy of Sciences, Kunming 650216, People's Republic of China}

\affiliation{International Center of Supernovae at the Yunnan Key Laboratory, Kunming 650216, People's Republic of China}
\affiliation{Center for Astronomical Mega-Science, Chinese Academy of Sciences, Beijing, 100012, People's Republic of China}

\author{Tao Wu}
\affiliation{Yunnan Observatories, Chinese Academy of Sciences, Kunming 650216, People's Republic of China}
\affiliation{International Center of Supernovae at the Yunnan Key Laboratory, Kunming 650216, People's Republic of China}
\affiliation{University of Chinese Academy of Sciences, Beijing 100049, People's Republic of China}
\affiliation{Center for Astronomical Mega-Science, Chinese Academy of Sciences, Beijing, 100012, People's Republic of China}

\author{Yuetong Wang}
\affiliation{Yunnan Observatories, Chinese Academy of Sciences, Kunming 650216, People's Republic of China}

\affiliation{International Center of Supernovae at the Yunnan Key Laboratory, Kunming 650216, People's Republic of China}
\affiliation{University of Chinese Academy of Sciences, Beijing 100049, People's Republic of China}



\begin{abstract}
The magnetic field is a significant and universal physical phenomenon in modern astrophysics. Small-scale magnetic fields are very important in the stellar atmosphere. They are ubiquitous, and strongly couple with the acoustic waves. Therefore, their presence affects the properties of acoustic waves in the stellar outer layer. In the present work, under the assumption that the small-scale magnetic features are the cause of the asteroseismic surface term (the frequency-dependent frequency offset between stars and their models), we explore the strength of such fields in the solar analog KIC 8006161. By considering the effect of small-scale magnetic fields in the stellar photosphere, we use the observed oscillation frequencies to constrain the inner structures and surface small-scale magnetic fields of solar-like star KIC 8006161. To agree with the existing observations, such as oscillation frequencies, and their frequency separation ratios,
 the theoretical model requires a small-scale magnetic field to form a magnetic-arch splicing layer in the stellar
 outer atmosphere.
  The small-scale magnetic field strengths for KIC 8006161 obtained from best-fit model with $Y_{\rm init} = 0.249+1.33Z_{\rm init}$ and $Y_{\rm init}$ as a free parameter, are approximately 96 G and 89 G, respectively. The corresponding locations of the magnetic-arch splicing layer are about $522$ km and 510 km, respectively.

\end{abstract}



\section{Introduction}

The small-scale magnetic fields in the quiet solar photosphere are a fundamental component of solar magnetism. They are ubiquitous on the solar surface, and store a substantial mount of magnetic energy (e.g. \citealp{1994ASSL..189.....S}; \citealp{2004Natur.430..326T}). They may contribute to coronal heating and the acceleration of the stellar wind (\citealp{1998Natur.394..152S}; \citealp{2003ApJ...597L.165S}). Comprehending the small-scale magnetic fields may help us to solve many of the key problems in solar and stellar physics.

Currently, an increasing amount of observational data with high spatial resolution and polarization sensitivity is used to investigate the small-scale magnetic fields of the solar photosphere. \cite{1996ApJ...460.1019L} first discovered the isolated, small-scale horizontal magnetic fields using the observations from the Advanced Stokes Polarimeter. \cite{2004Natur.430..326T} used observed Hanle depolarization and a three-dimensional radiative transfer model to indicate that there was an ubiquitous tangled magnetic field with a strength of approximately 130 G. Based on observations of the Solar Optical Telescope onboard the Hinode space observatory with very high spatial resolution, \cite{2008ApJ...672.1237L}  obtained a mean field strength of about 55 G for horizontal fields, and around 11 G for vertical magnetic fields. They found the vertical fields were concentrated in the intergranular lanes, while the horizontal fields were mostly located at the edges of the bright granules. \cite{2015ApJ...807...70J} used the spectro-polarimeter observations on Hinode to analyze the cyclic variation of the inter-network horizontal fields. They found that the horizontal fields with about 87 G did not vary with solar activity. Many scholars have also used a amount of high-precision spectro-polarimetric data to probe horizontal fields in the quiet photosphere(i.e. \citealp{2010ApJ...723L.149D}; \citealp{2016A&A...596A...5M}; \citealp{2017ApJ...835...14L}). Furthermore, \cite{2020A&A...642A.210M} used a spatially-regularized weak-field approximation (WFA) method to derive the stratification of the magnetic field vector in plage regions.

In the numerical simulation, \citet{2005ESASP.596E..65S,2006ASPC..354..345S} utilized the $\rm CO^5BOLD$ code to carry out a three-dimensional magnetohydrodynamic simulations spanning from the convection zone to the chromosphere. They discovered a small-scale 'magnetic canopy' at an attitude of about 400 km. There was 'magnetic void' with a field strength below the canopy of less than 3 G. \cite{2007ApJ...665.1469A} simulated the quiet sun from the upper convection zone to the lower corona. They found long, horizontal ribbons of magnetic flux threading through the low chromosphere. \cite{2008A&A...481L...5S} carried out the radiative MHD simulations of dynamo action driven by near-surface convection and got that horizontal fields predominated in the spectral lines observed by the Hinode spectro-polarimeter. The corresponding ratio of the horizontal and vertical fields was 4 to 6 which was consistent with the observed one (\citealp{2008ApJ...672.1237L}). \cite{2008ApJ...680L..85S} found that the horizontal field strength reached a local maximum at approximately 500 km in the photosphere, where the horizontal fields were 2 or 5.6 times stronger than the vertical component depending on the initial state or boundary conditions. Very similar results were also reported by \cite{2014ApJ...789..132R}.

 The small-scale magnetic fields in the photosphere are coupled with the acoustic waves(\citealp{2002ApJ...564..508R}; \citealp{2003ApJ...599..626B}; \citealp{2007AN....328..286C}). The ratio between magnetic pressure and gas pressure, although initially small, exhibits an exponential increase with altitude. This would affect the wave propagation speed. \cite{2021ApJ...916..107L} considered the effects of small-scale magnetic fields in the quiet solar photosphere on the frequencies of solar p-mode oscillation. They found the reflection of the solar p-mode at the magnetic-arch splicing layer. They got the solar small-scale magnetic field strength about 90 G and the corresponding location was at a height of about 630 km.

 For solar-like stars, there is a convective envelope on the stellar surface, which is very similar to that of the sun. Although, the origin of small-scale magnetic fields is still debated. Based on the theory of fast dynamo and numerical simulations, the thermally-driven turbulent convection was indicated to be an effective source for small-scale magnetic fields \citep{1999ApJ...515L..39C}. The convective envelope may make the small-scale magnetic fields widespread on the surface of solar-like stars. At present, the small-scale magnetic fields of the sun have been directly observed by the Hinode spectro-polarimeter. For solar-like stars at great distance, it is very difficult to directly observe the small-scale magnetic fields due to the spatial resolution. Because the small-scale magnetic fields are coupled with sound waves, this provides an opportunity to detect small-scale magnetic fields by astroseismology. By utilizing only the high-precision oscillation frequencies obtained from the Kepler short-cadence data (\citealp{2017ApJ...850..110L}), we can use asteroseismology to constrain the stellar interior structure and the small-scale magnetic fields for solar-like stars.

Because the magnetic fields can result in shifts in the oscillation frequencies (\citealp{1990MNRAS.242...25G}), asteroseismology can be used to detect magnetic fields by the shifts in the oscillation frequencies. \cite{2022Natur.610...43L} detected the asymmetries in the multiplets of 3 red giant stars observed by Kepler, and used the asteroseismology to measure magnetic fields in their cores. \cite{2023A&A...670L..16D} also detected the strong magnetic fields in the cores of 11 red giant stars by the g-mode period spacing.

In this work, we assume that the near-surface effect for solar-like stars is caused by the influence of the small-scale magnetic fields in the photosphere. We employ a method proposed by \cite{2021ApJ...916..107L} to explore the strength of such field in solar-like stars. We choose a solar analog KIC 8006161. KIC 8006161 (HD 173701) is a solar-like G dwarf with a $\sim 7.4$ yr activity cycle (\citealp{2018ApJ...852...46K}). Using the latest version of the Asteroseismic Modeling Portal(AMP), \cite{2017A&A...601A..67C} presented the basic parameters (i.e. mass, radius and age) of KIC 8006161, which were very similar to those of the sun except for a higher surface metallicity. In addition, using Kepler photometry, \cite{2018ApJ...852...46K} obtained a mean rotation period of 27 days for KIC 8006161 which is also very similar to that of the sun. Due to the possibility of comparing the stellar parameters and interior structure of KIC 8006161 with those of the sun, KIC 8006161 is considered the ideal target for studying small-scale magnetic fields by asteroseismology. Considering the effects of small-scale magnetic fields in the photosphere, we use the observed p-mode frequencies to constrain the small-scale magnetic field strength for KIC 8006161. In $\S$2, we present the stellar model with small-scale magnetic fields in the stellar photosphere. In $\S$3, we will show the results of the model and present the magnetic field strength for KIC 8006161. At last, we'll give our summaries.

\section{Stellar model with small-scale magnetic fields}
We use the MESA evolutionary code with version r-10398 to construct stellar structure and evolutionary models (e.g. \citealp{2011ApJS..192....3P,2018ApJS..234...34P}). The OPAL equation of state tables are taken from \cite{2002ApJ...576.1064R} and the OPAL opacity tables are obtained from \cite{1996ApJ...464..943I}. For simplicity, we adopt the Eddington gray-atmosphere model as the stellar atmosphere model and use the GS98 (\citealp{1998SSRv...85..161G}) chemical composition. In the stellar structure and evolutionary models, we treat convection as the standard mixing length theory (\citealp{1958ZA.....46..108B}), and do not consider convective overshooting. We use the default parameters to include the element diffusion. In addition, we also take into account the radiative diffusivity (\citealp{2002A&A...390..611M}). In order to get the theoretical p-mode oscillation frequencies which are compared with the observed ones, we use the "adipls" package (\citealp{2008Ap&SS.316..113C}) in MESA to solve the adiabatic oscillation equations. In the evolutionary models, in order to get more detailed information about stellar structure, we set the outer boundary of the stellar atmosphere with a very low density $10^{-10}$ and adjust the mesh with at least 8000 cells.

\cite{2021ApJ...916..107L} modeled the effect of small-scale magnetic fields on solar oscillation frequencies by adding magnetic pressure in the solar atmosphere model. Based on their work, we construct the stellar interior structure of the solar analog KIC 8006161 with the effects of magnetic fields in the photosphere. In the Eddington gray-atmosphere model, we use the unified $T-\tau$ relation (\citealp{2021ApJ...916..107L}):
\begin{equation}
T^{4}=\frac{3}{4}T_{\rm eff}^{4}[\tau+q(\tau)]
\end{equation}
In the above equation, an additional term in the Hopf-like function $q(\tau)$ is added: $q(\tau) = q_{\rm ori}(\tau)+a \exp (-b\sqrt{\tau})$, in which the second term mimics the effect of magnetic fields. In the Eddington gray-atmosphere model, $q_{\rm ori}$ is $2/3$. Parameters $a$ and $b$ represent the magnetic field strength and the location of the magnetic-arch splicing layer, respectively. In Eq. (1), the additional term indicates an increase in temperature in the photosphere. The part where the radiative pressure increases represents the effect of the magnetic pressure. In the stellar photosphere, as the magnetic pressure increases, the gas pressure decreases which leads to a rapid decrease in density. The sound speed may increase rapidly. When the acoustic waves propagate to the magnetic-arch splicing layer, they will be reflected or transmitted. For the reflected standing waves, $P^{'}=0$ is adopted as the surface mechanical boundary condition. For more details, please refer to \cite{2021ApJ...916..107L}.

 By using the Kepler short-cadence data, \cite{2017ApJ...850..110L} extracted 54 oscillation frequencies for KIC 8006161 and presented the corresponding frequency ratios. The errors of the observed frequencies and frequency ratios are also presented in \cite{2017ApJ...850..110L}. Using the Kepler observed 54 modes with angular degree $l=0\sim3$, we construct the stellar interior structure for KIC 8006161 with magnetic fields in the photosphere.

Based on the spectroscopic observations, we set wide ranges of initial input parameters for the theoretical models: $M\sim 0.96-1.03M_{\odot}$ with a step of $0.005M_{\odot}$, $Z_{\rm init} \sim 0.03-0.04$ with a step of $0.0005$, $\alpha \sim 1.7-2.32$ with a step of $0.04$. We use the relation of $Y_{\rm init}= 0.249+1.33Z_{\rm init}$ (\citealp{2018MNRAS.475..981L}) to calculate the initial helium abundance.

\section{Calculations and results}
\subsection{Theoretical results}

\begin{figure}
\gridline{\fig{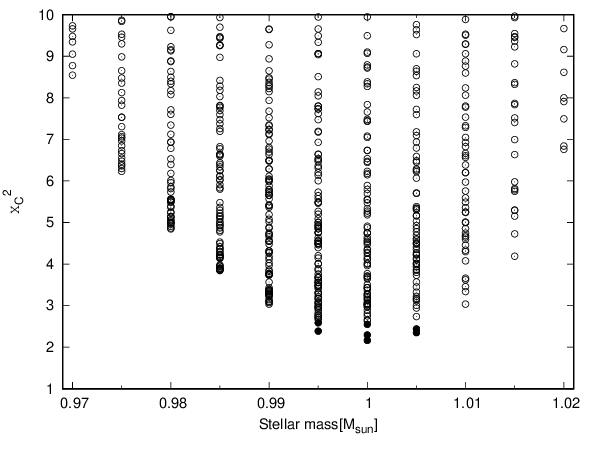}{0.33\textwidth}{}
          \fig{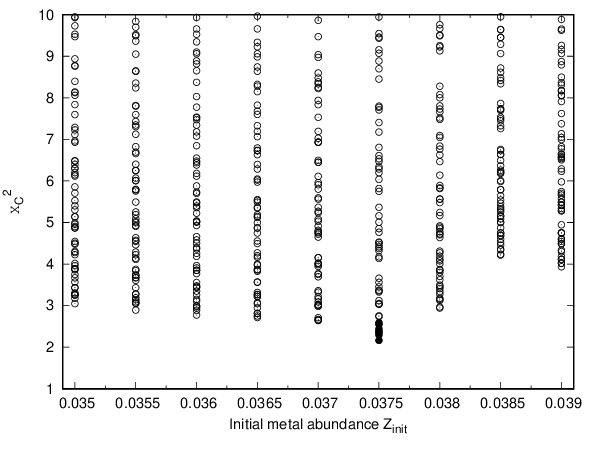}{0.33\textwidth}{}
          \fig{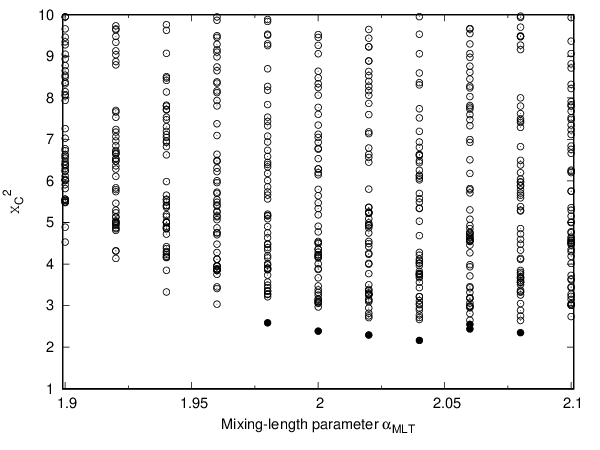}{0.33\textwidth}{}
           }

\caption{The distribution of $\chi^{2}_{\rm C}$ with stellar parameters. The panels from left to right represent the distribution of $\chi^{2}_{\rm C}$ with stellar mass, initial metal abundance $Z_{\rm init}$, and mixing-length parameter $\alpha_{\rm MLT}$, respectively. The filled circles represent the results of models with $\chi_{\rm C}^{2} < 2.6$. \label{fig:initial_model}}
\end{figure}

\begin{figure}[ht!]
\plotone{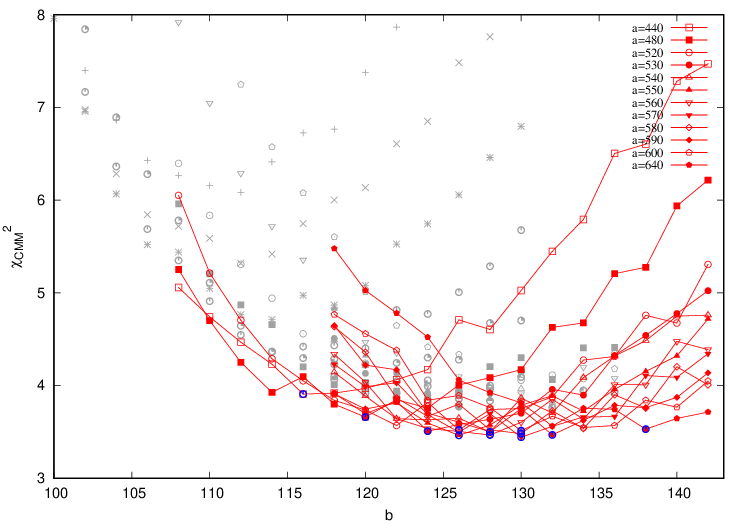}
\caption{The distribution of $\chi_{\rm CMM}^{2}$ with the parameters $a$ and $b$ of magnetic fields. The grey points represent the results obtained from the model with initial fundamental parameters, while the red lines correspond to the results derived from the $\chi^{2}$-minimization model with $M=1.0M_{\odot}$, $Z_{\rm init} = 0.032$, $\alpha = 2.12$. The points sharing the same type indicate the results with an identical parameter $a$. \label{fig:logXi2_b_a}}
\end{figure}
In our work, based on \cite{2021ApJ...916..107L}, we introduce the magnetic pressure in the stellar photosphere. Due to the coupling between the magnetic fields and the acoustic waves, the magnetic field strength and the location of the magnetic-arch splicing layer in the photosphere must be very important for stellar models.

 In order to reduce the amount of calculations, we use an iterative computation method to get self-consistent magnetic parameters $a$ and $b$. The small-scale magnetic fields in the stellar photosphere only influent the stellar surface structure. For solar-like stars, \cite{2003A&A...411..215R} found that the ratios $r_{\rm ij}$ of small to large-frequency separations are only determined by the stellar interior structure. The ratio of small to large-frequency separations $r_{\rm ij}$ are defined as follows (\citealp{2003A&A...411..215R}):
 \begin{equation}
 r_{01}(n) = \frac{\delta \nu_{01}(n)}{\Delta\nu_{l=1}(n)}, r_{02}(n) = \frac{\delta\nu_{02}(n)}{\Delta\nu_{l=1}(n)}
 \end{equation}
 and
 \begin{equation}
 r_{10}(n) = \frac{\delta\nu_{10}(n)}{\Delta\nu_{l=0}(n+1)}
 \end{equation}
 in which
 \begin{equation}
 \Delta\nu_{l}(n) = \nu_{n,l}-\nu_{n-1,l}
 \end{equation}
 \begin{equation}
 \delta\nu_{02}(n) = \nu_{n,0}-\nu_{n-1,2}
 \end{equation}
 \begin{equation}
 \delta\nu_{01}(n) = \frac{1}{8}(\nu_{n-1,0}-4\nu_{n-1,1}+6\nu_{n,0}-4\nu_{n,1}+\nu_{n+1,0})
 \end{equation}
 and
 \begin{equation}
 \delta\nu_{10}(n)=-\frac{1}{8}(\nu_{n-1,1}-4\nu_{n,0}+6\nu_{n,1}-4\nu_{n+1,0}+\nu_{n+1,1}).
 \end{equation}

 This means that if the stellar interior structure remains unchanged, changing only the surface structure will not change the ratios $r_{\rm ij}$. For KIC 8006161, we use the observed frequency ratios $r_{\rm ij}$ and the spectral parameters (i.e. $T_{\rm eff}$, $\log g$, $[\rm Fe/H]$) to constrain the stellar initial fundamental parameters. We use a $\chi_{C}^{2}$ minimization model as an initial one through a comparison of models with observations:
\begin{equation}
\chi_{C}^{2} = \sum_{i=1}^{4}(\frac{C_{i}^{\rm theo}-C_{i}^{\rm obs}}{\sigma_{ C_{i}^{\rm obs}}})^{2}
\end{equation}
in which $C = (r_{\rm ij},T_{\rm eff}, \log g, [\rm Fe/H])$. $C_{i}^{\rm theo}$ is the theoretical values and $C_{i}^{\rm obs}$ is the observational ones. $\sigma _{C_{i}^{\rm obs}}$ is the errors of the observations. The distribution of $\chi_{C}^{2}$ with the parameters is shown in Fig. \ref{fig:initial_model}. We choose the model with the minimum $\chi_{C}^{2}$ as the initial fundamental model:  $M=1.0M_{\odot}$, $Z_{\rm init} = 0.0375$, $\alpha = 2.04$.

Using the initial fundamental parameters, we preliminarily determine the values of $a$ and $b$. Then, we calculate the stellar structure and evolution of the models with these values, and get the $\chi^{2}$-minimization model. In this model, it is defined as (i.e. \citealp{2004A&A...417..235E}, \citealp{2016ApJ...818L..13W})
\begin{equation}
\chi_{\rm CMM}^{2} = \frac{1}{N}\sum_{i=1}^{N}(\frac{\nu_{i}^{\rm obs}-\nu_{i}^{\rm theo}}{\sigma_{\nu_{i}^{\rm obs}}})^{2}
\end{equation}
in which $\nu_{i}^{\rm obs}$ and $\nu_{i}^{\rm theo}$ are observed frequencies and theoretical ones, respectively. $\sigma_{\nu_{i}^{\rm obs}}$ is the errors of the observed frequencies. N is the total observed frequencies number. When the initial input fundamental parameters are the same as those of the $\chi^{2}$-minimization model, the values of $a$ and $b$ are adopted as the optimal ones. If they are different, the aforementioned process must be iterated until the condition is satisfied. We also explore the utilization of diverse initial fundamental parameters. After iterative computation, the optimal values of parameters $a$ and $b$ exhibit remarkable proximity.

   \begin{figure}
\gridline{\fig{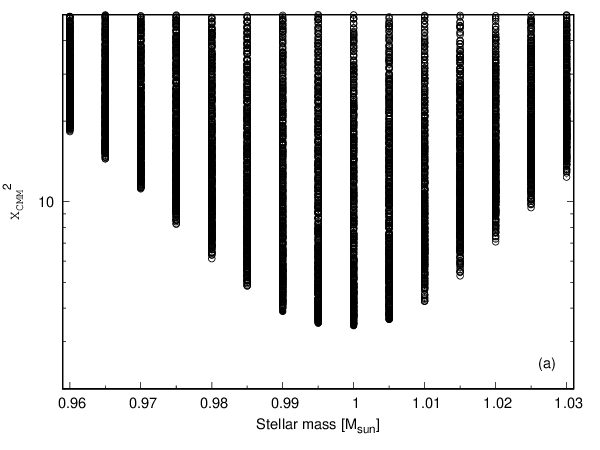}{0.33\textwidth}{}
          \fig{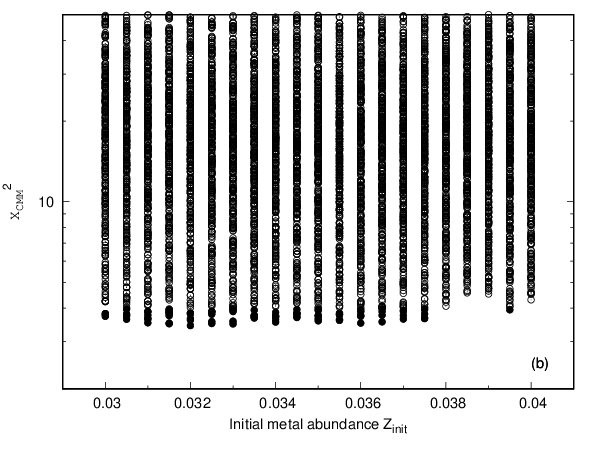}{0.33\textwidth}{}
          \fig{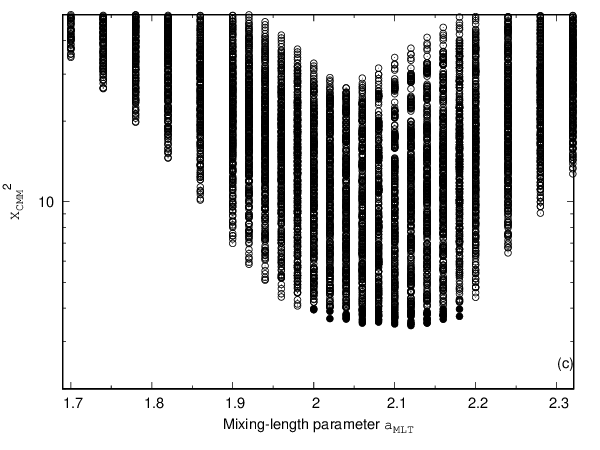}{0.33\textwidth}{}
           }
\gridline{\fig{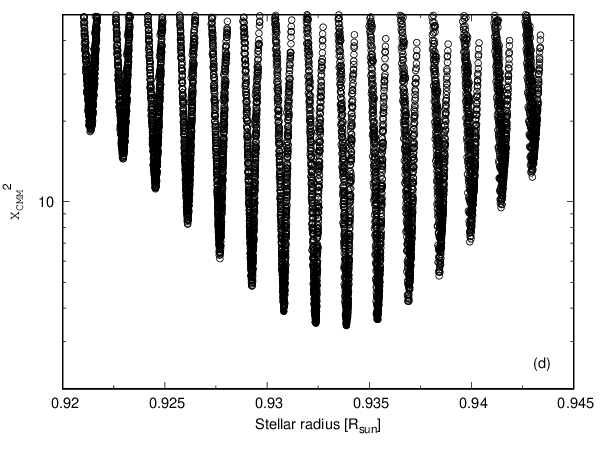}{0.33\textwidth}{}
           \fig{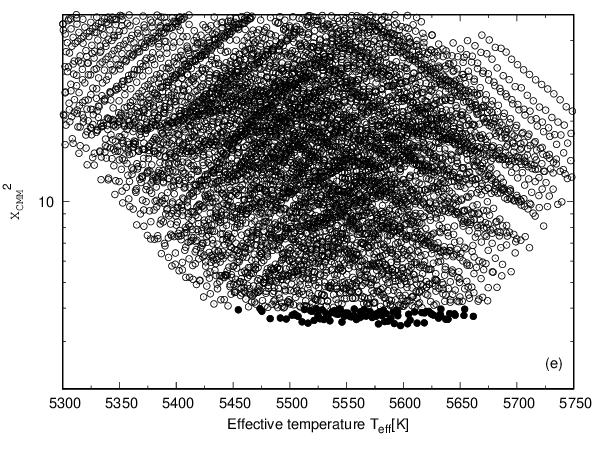}{0.33\textwidth}{}
           \fig{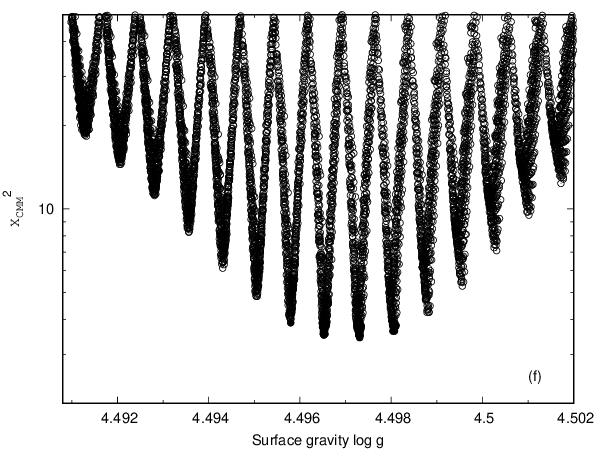}{0.33\textwidth}{}
           }
\gridline{\fig{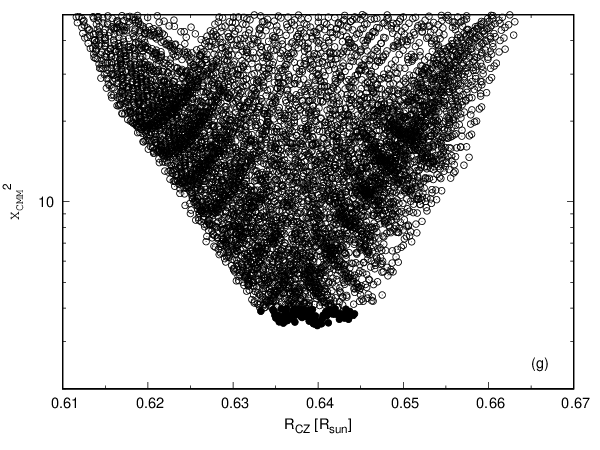}{0.33\textwidth}{}
          \fig{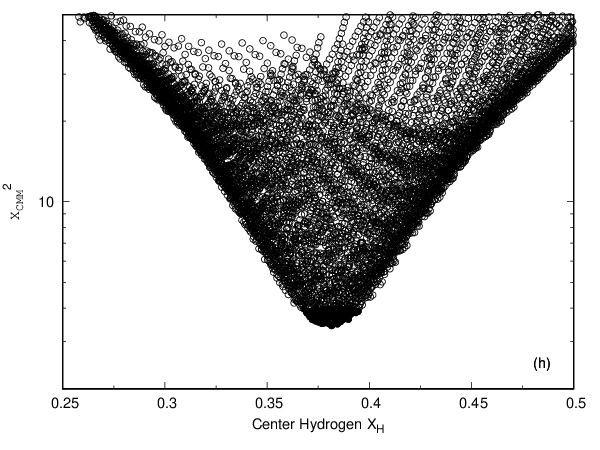}{0.33\textwidth}{}
           \fig{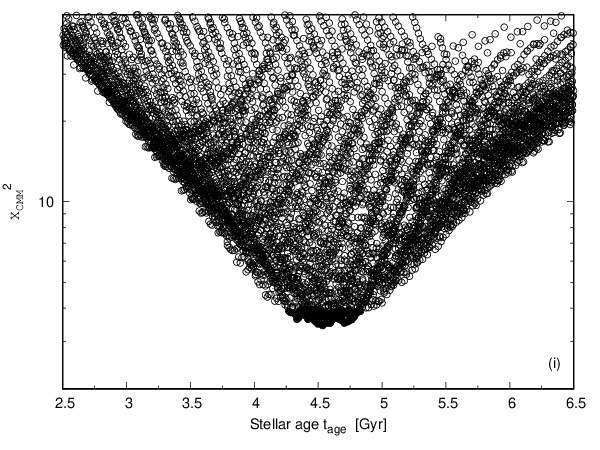}{0.33\textwidth}{}
         }
\caption{The distribution of $\chi^{2}_{\rm CMM}$ with stellar parameters when fixing the magnetic parameters as $a=570$ and $b = 130$ and setting the initial helium abundance as $Y_{\rm init}=0.249+1.33Z_{\rm init}$. The panels (a)-(i) represent the distribution of $\chi^{2}_{\rm CMM}$ with stellar mass, initial metal abundance $Z_{\rm init}$, mixing-length parameter $\alpha_{\rm MLT}$, stellar radius, effective temperature $T_{\rm eff}$, surface gravity $\log g$, location of convection zone bottom $R_{\rm CZ}$,  center hydrogen $X_{\rm H}$, and stellar age $t_{\rm age}$, respectively. The filled circles represent the results of better candidate models in Table \ref{tbl-candidate_model}. \label{fig:bestfit_model}}
\end{figure}

According to the work of \cite{2016ApJ...818L..13W}, there is one $\chi^{2}$-minimization model (CMM) for an evolutionary track with a set of fundamental parameters. The optimal magnetic parameters $a$ and $b$ for KIC 8006161 are 570 and 130, respectively, after two iterations. Fig. \ref{fig:logXi2_b_a} presents the distribution of $\chi_{\rm CMM}^{2}$ with the parameters $a$ and $b$ of magnetic fields. The grey points are the distribution of $\chi_{\rm CMM}^{2}$ for the model with initial fundamental parameters. The red points are those obtained from the $\chi$-minimization model with $M=1.0M_{\odot}$, $Z_{\rm init} = 0.032$ and $\alpha = 2.12$ after 2 iterations. It can be seen that when the value of $a$ is fixed and $b$ increases, the value of $\chi_{\rm CMM}^{2}$ firstly decreases, then increases. The positions with lowest $\chi_{\rm CMM}^{2}$ for different $a$ are marked with blue dots. As the parameter $a$ increases, $\chi_{\rm CMM}^{2}$ gradually decreases until it reaches a minimum at $a=570$, after which it gradually increases. Using the iterative calculations method described above, we get the fundamental parameters of the model with the minimum $\chi^{2}_{\rm CMM}=3.444$: $M=1.0M_{\odot}$, $Z_{\rm init} = 0.032$, $\alpha = 2.12$. We use the models with $\chi_{\rm CMM}^{2} < 4.0$ (i.e. within twice the error bar of observed frequencies) as better candidate models, and their fundamental parameters are listed in Table \ref{tbl-candidate_model}. Among these candidate models, we select model M48 with a minimum $\chi_{\rm CMM}^{2}=3.44$ as the best-fit model. We use the notation $x_{-dx}^{+dx}$ to describe the values of stellar parameters, where $x$ represents the value of the best-fit model, and $dx$ represents the maximum deviation between the candidate models and the best-fit model. The corresponding stellar parameters are listed in Table \ref{tbl-parameter_compare}.

\begin{center}
\setlength{\LTcapwidth}{18cm}
\begin{longtable}{|c|c|c|c|c|c|c|c|c|c|c|c|c|c|c|}
%\captionsetup[longtable]{singlelinecheck=off} \\
\caption{The fundamental parameters for the better candidate models with the parameters a=570, b=130 and $\chi_{\rm CMM}^{2} < 4.0$.\label{tbl-candidate_model}}\\
\hline
No.  &  M[$M_{\odot}$]  & $Z_{\rm init}$ &  $\alpha_{\rm MLT}$  &  R[$R_{\odot}$]   &  $T_{\rm eff}$  &   L[$L_{\odot}$]   &   $\log g$  &  $X_{\rm H}$ &
 $t_{\rm age}$[Gyr]  &  $R_{\rm CZ}$[$R_{\odot}$]        &  [Fe/H]  &    $X_{s}$  &   $Y_{s}$  &  $\chi_{\rm CMM}^{2}$  \\

\hline
\endfirsthead % 表格第一页的表头
\hline
No.  &  M[$M_{\odot}$]  & $Z_{\rm init}$ &  $\alpha_{\rm MLT}$  &  R[$R_{\odot}$]   &  $T_{\rm eff}$  &   L[$L_{\odot}$]   &   $\log g$  &  $X_{\rm H}$ &
 $t_{\rm age}$[Gyr]  &  $R_{\rm CZ}$[$R_{\odot}$]        &  [Fe/H]  &    $X_{s}$  &   $Y_{s}$  &  $\chi_{\rm CMM}^{2}$  \\
%      & [$M_{\odot}$] &        &                      &  [$R_{\odot}$]&        &  [$L_{\odot}$]&     &               &
%[Gyr]           &  [$R_{\odot}$]        &          &             &            &                \\
\hline
\endhead % 表格接下来的每页的表头
\hline
\endfoot % 表格每页的表尾
\hline
\endlastfoot % 表格的最后一页的表尾
\hline
\endlastfoot

% 示例数据
\hline
1      &0.990    &0.0370    &2.02     &0.931     &5475    &0.700   &4.496   &0.375   &4.830  &0.633   &0.339   &0.689   &0.276   &3.895  \\
2      &0.995    &0.0315    &2.10     &0.932     &5583    &0.758   &4.497   &0.378   &4.685  &0.639   &0.261   &0.701   &0.269   &3.940  \\
3      &0.995    &0.0315    &2.12     &0.932     &5591    &0.763   &4.497   &0.372   &4.762  &0.638   &0.261   &0.702   &0.269   &3.867  \\
4      &0.995    &0.0320    &2.10     &0.932     &5576    &0.755   &4.497   &0.377   &4.700  &0.638   &0.269   &0.700   &0.270   &3.814 \\
5      &0.995    &0.0320    &2.12     &0.932     &5585    &0.760   &4.497   &0.371   &4.776  &0.637   &0.268   &0.700   &0.270   &3.820  \\
6      &0.995    &0.0325    &2.08     &0.932     &5562    &0.747   &4.497   &0.381   &4.641  &0.638   &0.277   &0.699   &0.271   &3.881  \\
7      &0.995    &0.0325    &2.10     &0.932     &5571    &0.752   &4.497   &0.375   &4.720  &0.637   &0.276   &0.699   &0.271   &3.883  \\
8      &0.995    &0.0325    &2.12     &0.932     &5579    &0.757   &4.497   &0.370   &4.795  &0.637   &0.276   &0.699   &0.270   &3.886  \\
9      &0.995    &0.0330    &2.08     &0.932     &5557    &0.745   &4.497   &0.379   &4.662  &0.637   &0.284   &0.698   &0.271   &3.973  \\
10     &0.995    &0.0330    &2.10     &0.932     &5565    &0.749   &4.497   &0.374   &4.740  &0.637   &0.284   &0.698   &0.271   &3.953  \\
11     &0.995    &0.0335    &2.08     &0.932     &5551    &0.742   &4.497   &0.378   &4.682  &0.637   &0.291   &0.697   &0.272   &3.760  \\
12     &0.995    &0.0335    &2.10     &0.932     &5560    &0.746   &4.497   &0.372   &4.760  &0.637   &0.291   &0.697   &0.272   &3.675 \\
13     &0.995    &0.0340    &2.06     &0.932     &5537    &0.734   &4.497   &0.382   &4.622  &0.638   &0.299   &0.695   &0.273   &3.855  \\
14     &0.995    &0.0340    &2.08     &0.932     &5546    &0.739   &4.497   &0.377   &4.703  &0.637   &0.299   &0.695   &0.273   &3.550   \\
15     &0.995    &0.0340    &2.10     &0.932     &5555    &0.743   &4.497   &0.371   &4.782  &0.637   &0.298   &0.696   &0.272   &3.715 \\
16     &0.995    &0.0345    &2.06     &0.932     &5532    &0.731   &4.497   &0.381   &4.642  &0.637   &0.306   &0.694   &0.274   &3.819  \\
17     &0.995    &0.0345    &2.08     &0.932     &5541    &0.736   &4.497   &0.375   &4.724  &0.636   &0.305   &0.694   &0.273   &3.812 \\
18     &0.995    &0.0345    &2.10     &0.932     &5549    &0.741   &4.497   &0.370   &4.803  &0.636   &0.305   &0.695   &0.273   &3.836  \\
19     &0.995    &0.0350    &2.06     &0.932     &5527    &0.728   &4.497   &0.380   &4.664  &0.637   &0.313   &0.693   &0.274   &3.652 \\
20     &0.995    &0.0350    &2.08     &0.932     &5535    &0.733   &4.497   &0.374   &4.744  &0.636   &0.312   &0.693   &0.274   &3.586    \\
21     &0.995    &0.0350    &2.10     &0.932     &5544    &0.738   &4.496   &0.368   &4.823  &0.636   &0.312   &0.694   &0.274   &3.977  \\
22     &0.995    &0.0355    &2.06     & 0.932    &5521    &0.726   &4.497   &0.378   &4.684  &0.636   &0.320   &0.692   &0.275   &3.599     \\
23     &0.995    &0.0355    &2.08     &0.932     &5530    &0.730   &4.497   &0.372   &4.766  &0.636   &0.319   &0.692   &0.275   &3.630 \\
24     &0.995    &0.0360    &2.04     &0.932     &5507    &0.718   &4.497   &0.382   &4.621  &0.637   &0.327   &0.691   &0.276   &3.730 \\
25     &0.995    &0.0360    &2.06     &0.932     &5516    &0.723   &4.497   &0.377   &4.704  &0.636   &0.327   &0.691   &0.275   &3.519    \\
26     &0.995    &0.0360    &2.08     &0.932     &5525    &0.728   &4.496   &0.371   &4.787  &0.635   &0.326   &0.691   &0.275   &3.766  \\
27     &0.995    &0.0365    &2.04     &0.932     &5502    &0.716   &4.497   &0.381   &4.643  &0.636   &0.334   &0.689   &0.276   &3.765  \\
28     &0.995    &0.0365    &2.06     &0.932     &5511    &0.720   &4.497   &0.375   &4.726  &0.635   &0.333   &0.690   &0.276   &3.557  \\
29     &0.995    &0.0365    &2.08     &0.932     &5519    &0.725   &4.496   &0.369   &4.807  &0.635   &0.333   &0.690   &0.276   &3.888  \\
30     &0.995    &0.0370    &2.04     &0.932     &5497    &0.713   &4.497   &0.379   &4.663  &0.636   &0.340   &0.688   &0.277   &3.642 \\
31     &0.995    &0.0370    &2.06     &0.932     &5506    &0.718   &4.496   &0.374   &4.747  &0.635   &0.340   &0.689   &0.277   &3.760  \\
32     &0.995    &0.0375    &2.00     &0.932     &5474    &0.701   &4.497   &0.390   &4.510  &0.636   &0.348   &0.687   &0.278   &3.984  \\
33     &0.995    &0.0375    &2.02     &0.932     &5483    &0.706   &4.497   &0.384   &4.598  &0.636   &0.348   &0.687   &0.278   &3.655 \\
34     &0.995    &0.0375    &2.04     &0.932     &5492    &0.710   &4.497   &0.378   &4.683  &0.635   &0.347   &0.687   &0.278   &3.675 \\
35     &0.995    &0.0395    &2.00     &0.932     &5455    &0.691   &4.497   &0.380   &4.670  &0.635   &0.373   &0.683   &0.280   &3.945  \\

36      &1.000    &0.0300    &2.14     &0.934     &5632    &0.789   &4.497   &0.380   &4.551  &0.642   &0.238   &0.704   &0.268   &3.801  \\
37      &1.000    &0.0300    &2.16     &0.934     &5641    &0.793   &4.497   &0.374   &4.625  &0.641   &0.238   &0.705   &0.267   &3.821  \\
38      &1.000    &0.0305    &2.14     &0.934     &5625    &0.784   &4.497   &0.379   &4.568  &0.642   &0.246   &0.703   &0.268   &3.805  \\
39      &1.000    &0.0305    &2.16     &0.934     &5633    &0.789   &4.497   &0.373   &4.641  &0.641   &0.246   &0.704   &0.268   &3.734  \\
40      &1.000    &0.0310    &2.12     &0.934     &5610    &0.776   &4.497   &0.384   &4.507  &0.641   &0.255   &0.702   &0.269   &3.874  \\
41      &1.000    &0.0310    &2.14     &0.934     &5618    &0.781   &4.497   &0.378   &4.582  &0.641   &0.254   &0.702   &0.269   &3.527  \\
42      &1.000    &0.0310    &2.16     &0.934     &5627    &0.785   &4.497   &0.373   &4.655  &0.641   &0.254   &0.702   &0.269   &3.766  \\
43      &1.000    &0.0315    &2.10     &0.934     &5595    &0.768   &4.497   &0.388   &4.444  &0.641   &0.263   &0.700   &0.270   &3.749  \\
44      &1.000    &0.0315    &2.12     &0.934     &5604    &0.773   &4.497   &0.383   &4.520  &0.640   &0.262   &0.701   &0.270   &3.500   \\
45      &1.000    &0.0315    &2.14     &0.934     &5612    &0.777   &4.497   & 0.377  &4.595  &0.640   &0.262   &0.701   &0.270   &3.519     \\
46      &1.000    &0.0315    &2.16     &0.934     &5620    &0.782   &4.497   &0.372   &4.669  &0.639   &0.261   &0.701   &0.269   &3.890 \\
47      &1.000    &0.0320    &2.10     &0.934     &5589    &0.764   &4.497   &0.387   &4.457  &0.641   &0.271   &0.699   &0.271   &3.659 \\
48      &1.000    &0.0320    &2.12     &0.934     &5598    &0.769   &4.497   &0.382   &4.535  &0.640   &0.270   &0.700   &0.270   &3.444     \\
49      &1.000    &0.0320    &2.14     &0.934     &5606    &0.774   &4.497   &0.376   &4.609  &0.639   &0.269   &0.700   &0.270   &3.609      \\
50      &1.000    &0.0325    &2.10     &0.934     &5584    &0.762   &4.497   &0.386   &4.477  &0.640   &0.278   &0.698   &0.271   &3.564      \\
51      &1.000    &0.0325    &2.12     &0.934     &5592    &0.766   &4.497   &0.380   &4.555  &0.639   &0.277   &0.698   &0.271   &3.505      \\
52      &1.000    &0.0330    &2.10     &0.934     &5578    &0.759   &4.497   &0.384   &4.498  &0.640   &0.285   &0.697   &0.272   &3.493     \\
53      &1.000    &0.0330    &2.12     &0.934     &5587    &0.763   &4.497   &0.379   &4.574  &0.640   &0.285   &0.697   &0.272   &3.580     \\
54      &1.000    &0.0330    &2.14     &0.934     &5595    &0.768   &4.497   &0.373   &4.650  &0.638   &0.284   &0.698   &0.271   &3.887  \\
55      &1.000    &0.0335    &2.08     &0.934     &5564    &0.751   &4.497   &0.389   &4.438  &0.640   &0.293   &0.696   &0.273   &3.742  \\
56      &1.000    &0.0335    &2.10     &0.934     &5573    &0.756   &4.497   &0.383   &4.518  &0.639   &0.292   &0.696   &0.273   &3.692 \\
57      &1.000    &0.0335    &2.12     &0.934     &5581    &0.760   &4.497   &0.377   &4.595  &0.638   &0.292   &0.696   &0.272   &3.753  \\
58      &1.000    &0.0340    &2.08     &0.934     &5559    &0.748   &4.497   &0.387   &4.459  &0.639   &0.300   &0.695   &0.273   &3.631 \\
59      &1.000    &0.0340    &2.10     &0.934     &5567    &0.753   &4.497   &0.381   &4.538  &0.639   &0.300   &0.695   &0.273   &3.737 \\
60      &1.000    &0.0340    &2.12     &0.934     &5576    &0.757   &4.497   &0.376   &4.615  &0.638   &0.299   &0.695   &0.273   &3.727 \\
61      &1.000    &0.0345    &2.06     &0.934     &5545    &0.741   &4.497   &0.392   &4.398  &0.639   &0.308   &0.693   &0.274   &3.957 \\
62      &1.000    &0.0345    &2.08     &0.934     &5553    &0.745   &4.497   &0.386   &4.480  &0.639   &0.307   &0.694   &0.274   &3.720 \\
63      &1.000    &0.0345    &2.10     &0.934     &5562    &0.750   &4.497   &0.380   &4.558  &0.638   &0.307   &0.694   &0.274   &3.692 \\
64      &1.000    &0.0350    &2.06     &0.934     &5539    &0.738   &4.497   &0.390   &4.418  &0.639   &0.315   &0.692   &0.275   &3.922 \\
65      &1.000    &0.0350    &2.08     &0.934     &5548    &0.742   &4.497   &0.384   &4.499  &0.638   &0.314   &0.692   &0.275   &3.731 \\
66      &1.000    &0.0350    &2.10     &0.934     &5557    &0.747   &4.497   &0.378   &4.579  &0.639   &0.313   &0.693   &0.274   &3.761  \\
67      &1.000    &0.0355    &2.06     &0.934     &5534    &0.735   &4.497   &0.389   &4.439  &0.639   &0.322   &0.691   &0.276   &3.779  \\
68      &1.000    &0.0355    &2.08     &0.934     &5543    &0.740   &4.497   &0.383   &4.519  &0.638   &0.321   &0.691   &0.275   &3.796  \\
69      &1.000    &0.0355    &2.10     &0.934     &5551    &0.744   &4.497   &0.377   &4.599  &0.638   &0.320   &0.692   &0.275   &3.834  \\
70      &1.000    &0.0360    &2.06     &0.934     &5529    &0.732   &4.497   &0.387   &4.459  &0.638   &0.328   &0.690   &0.276   &3.942 \\
71      &1.000    &0.0360    &2.08     &0.934     &5537    &0.737   &4.497   &0.381   &4.539  &0.638   &0.328   &0.690   &0.276   &3.883  \\
72      &1.000    &0.0365    &2.06     &0.934     &5523    &0.729   &4.497   &0.386   &4.478  &0.638   &0.335   &0.689   &0.277   &3.983 \\
73      &1.000    &0.0370    &2.06     &0.934     &5518    &0.727   &4.497   &0.384   &4.498  &0.638   &0.342   &0.688   &0.277   &3.920 \\
74      &1.000    &0.0375    &2.04     &0.934     &5504    &0.719   &4.497   &0.389   &4.436  &0.638   &0.349   &0.686   &0.278   &3.807  \\
75      &1.000    &0.0375    &2.06     &0.934     &5513    &0.724   &4.497   &0.383   &4.518  &0.637   &0.348   &0.687   &0.278   &3.956 \\

76      &1.005    &0.0300    &2.14     &0.935     &5645    &0.798   &4.498   &0.391   &4.317  &0.644   &0.240   &0.704   &0.268   &3.810 \\
77      &1.005    &0.0300    &2.16     &0.935     &5653    &0.803   &4.498   &0.385   &4.389  &0.644   &0.240   &0.704   &0.268   &3.759  \\
78      &1.005    &0.0300    &2.18     &0.935     &5662    &0.808   &4.498   &0.380   &4.461  &0.643   &0.239   &0.704   &0.268   &3.730 \\
79      &1.005    &0.0305    &2.14     &0.935     &5637    &0.794   &4.498   &0.390   &4.334  &0.644   &0.248   &0.702   &0.269   &3.635 \\
80      &1.005    &0.0305    &2.16     &0.935     &5646    &0.798   &4.498   &0.384   &4.406  &0.643   &0.247   &0.703   &0.269   &3.767 \\
81      &1.005    &0.0305    &2.18     &0.935     &5654    &0.803   &4.498   &0.379   &4.478  &0.643   &0.247   &0.703   &0.268   &3.965 \\
82      &1.005    &0.0310    &2.12     &0.935     &5623    &0.786   &4.498   &0.395   &4.271  &0.644   &0.256   &0.701   &0.270   &3.894 \\
83      &1.005    &0.0310    &2.14     &0.935     &5631    &0.790   &4.498   &0.389   &4.345  &0.643   &0.256   &0.701   &0.270   &3.671 \\
84      &1.005    &0.0310    &2.16     &0.935     &5639    &0.795   &4.498   &0.383   &4.419  &0.642   &0.255   &0.702   &0.269   &3.638 \\
85      &1.005    &0.0315    &2.12     &0.935     &5616    &0.782   &4.498   &0.393   &4.285  &0.643   &0.264   &0.700   &0.271   &3.850 \\
86      &1.005    &0.0315    &2.14     &0.935     &5625    &0.787   &4.498   &0.388   &4.359  &0.643   &0.264   &0.700   &0.270   &3.812 \\
87      &1.005    &0.0315    &2.16     &0.935     &5633    &0.792   &4.498   &0.382   &4.434  &0.642   &0.263   &0.700   &0.270   &3.969 \\
88      &1.005    &0.0320    &2.14     &0.935     &5619    &0.783   &4.498   &0.387   &4.374  &0.642   &0.271   &0.699   &0.271   &3.840 \\
89      &1.005    &0.0325    &2.12     &0.935     &5605    &0.776   &4.498   &0.391   &4.318  &0.642   &0.279   &0.698   &0.272   &3.797 \\
90      &1.005    &0.0335    &2.12     &0.935     &5594    &0.770   &4.498   &0.388   &4.357  &0.641   &0.294   &0.695   &0.273   &3.934 \\

% 继续添加更多行以使表格足够长，自动分页
\hline
\end{longtable}
\tablecomments{ The initial input parameters: stellar mass $M$, initial metal abundance $Z_{\rm init}$, and mixing length parameter $\alpha_{\rm MLT}$ are listed in column 2-4. The stellar parameters in column 5-14 present stellar radius R, effective temperature $T_{\rm eff}$, luminosity L, surface gravity $\log g$, center hydrogen $X_{\rm H}$, stellar age $t_{\rm age}$, location of convection zone bottom $R_{\rm CZ}$, [Fe/H], surface hydrogen abundance $X_{\rm s}$, and surface helium abundance $Y_{\rm s}$, respectively. The last column presents the matching goodness $\chi_{\rm CMM}^{2}$, which is calculated using Eq. (9). These results consider the effects of small-scale magnetic fields.}
\end{center}

The distributions of $\chi^{2}_{\rm CMM}$ with stellar parameters for the models with $a=570, b = 130$ are shown in Fig. \ref{fig:bestfit_model}. The filled circles represent the results of the better candidate models in Table \ref{tbl-candidate_model}. The panels (a)-(c) in Fig. \ref{fig:bestfit_model} are the distributions of $\chi_{\rm CMM}^{2}$ with initial input parameters. The results show that as the stellar mass and mixing-length parameter $\alpha_{\rm MLT}$ increase, $\chi_{\rm CMM}^{2}$ decreased, reaching a minimum at $M = 1.0M_{\odot}$ and $\alpha_{\rm MLT} = 2.12$, respectively, before  gradually increasing again. According to these results, the mixing-length parameter $\alpha = 2.12$ for KIC 8006161 is greater than that of the standard solar model. In general, for solar-like stars, the larger the convective mixing length parameter, the higher the convection energy transfer efficiency, which results in a higher effective temperature. This star has a smaller stellar radius.
From panel (b), $\chi_{\rm CMM}^{2}$ is insensitive to the initial metal abundance $Z_{\rm init}$. $Z_{\rm init}$ for the better candidate models in Table 1 occupies a wide range in parameter space.

The panels (d)-(i) in Fig. \ref{fig:bestfit_model} show the distributions of $\chi_{\rm CMM}^{2}$ with stellar parameters. We find that the distributions of $\chi_{\rm CMM}^{2}$ with stellar parameters converge to narrow ranges. It is very useful to accurately determine these stellar parameters. From panel (d), for KIC 8006161, the stellar radius is about $0.934R_{\odot}$, which is smaller than the solar radius. This result corresponds to larger mixing length parameter. On the other hand, the scaling relation for the large-frequency separation $\Delta\nu$ from \cite{1995A&A...293...87K} is:$\Delta \nu/\Delta\nu_{\odot}\approx (M/M_{\odot})^{1/2}(R/R_{\odot})^{-3/2}$. For a solar-like star with $1M_{\odot}$, the stellar radius is inversely proportional to the large-frequency separation. The large-frequency separation of KIC 8006161 is about $\rm 149.4 \mu Hz$ (\citealp{2017ApJ...850..110L}), which is greater than that of the sun with 134.9 $\rm \mu Hz$. KIC 8006161 has a smaller stellar radius. From this figure, we can get that KIC 8006161 is a main sequence star with a convective envelope covering $0.315R$. The stellar central hydrogen mass fraction is about $0.382$, and the stellar age is about $4.53$ Gyr.

 In the calculations above, we use the chemical enrichment law $\Delta Y/\Delta Z = 1.33$ (\citealp{2018MNRAS.475..981L})to determine the initial helium abundance. \cite{2017ApJ...835..173S} used seven independent pipelines to determine the stellar properties of 66 main-sequence targets. The four pipelines presented the distribution of initial helium abundance and initial metal abundance for all targets. However, the slopes of the function relating initial helium abundance and initial metal abundance differ among them. \cite{2019MNRAS.483.4678V} used the measured surface abundances in combination with the settling predicted by the stellar models to obtain a Galactic enrichment ratio of $\Delta Y/\Delta Z = 1.226\pm0.849$. In our work, we also consider models with initial helium abundance as a free parameter for $a=570$ and $b=130$. We set the initial helium abundance $Y_{\rm init}$ to range from 0.26 to 0.32, in steps of 0.005. The stellar mass M varies from 0.98 to 1.04 $M_{\odot}$  with a step of $0.01M_{\odot}$. The ranges for the other parameters remain unchanged from the previous calculations.

After calculating the stellar structure and evolution of all the models, we obtain the best-fit model with a minimum $\chi_{\rm CMM}^{2}$ of 2.68. The corresponding parameters are $M=1.02M_{\odot}$, $Y_{\rm init} = 0.270$, $Z_{\rm init}=0.037$, and $\alpha=1.98$. Compared to the results from models with $Y_{\rm init} = 0.249+1.33Z_{\rm init}$, the best-fit model with the initial helium abundance as a free parameter has a lower $\chi_{\rm CMM}^{2}$ and a larger stellar mass. The initial helium abundance $Y_{\rm init}$ for the best-fit model, where $Y_{\rm init}$ is treated as a free parameter, is lower than the value obtained using the relation $Y_{\rm init} = 0.249+1.33Z_{\rm init}$. This result is very similar to those from \cite{2017ApJ...835..173S}.

  For comparison, the stellar parameters of our theoretical models and those obtained from the literatures (\citealp{2017A&A...601A..67C}, \citealp{2017ApJ...835..173S}) for KIC 8006161 are all listed in Table \ref{tbl-parameter_compare}. The basic parameters of \cite{2017A&A...601A..67C} were generated by the latest version of Asteroseismic Modeling Portal (AMP). \cite{2017ApJ...835..173S} used seven codes by the different teams to calculate the stellar properties.
  We find that the stellar mass from \cite{2017A&A...601A..67C} is in complete agreement with the one we obtain using $Y_{\rm init}=0.249+1.33Z_{\rm init}$. In contrast, the stellar mass from \cite{2017ApJ...835..173S} is smaller, while the models where $Y_{\rm init}$ is treated as a free parameter yield the highest mass. The values of other stellar parameters (i.e. $R$, $\log g$, $R_{\rm CZ}/R$, $X_{C}$) show a high degree of similarity, with some overlap. Due to the smaller stellar masses reported by \cite{2017ApJ...835..173S}, the corresponding stellar ages are larger.
 Compared to the Gaia data for KIC 8006161 which has a luminosity of $0.680L_{\odot}$ (\citealp{2018A&A...616A...1G}), all models except our results with $Y_{\rm init}=0.249+1.33Z_{\rm init}$ are consistent with the observed data. Due to the lower metallicity of our models with $Y_{\rm init} = 0.249+1.33Z_{\rm init}$, the stellar luminosity and effective temperature are larger than those from other models. However, all of them overlap with the results of spectral parameters: $T_{\rm eff} = 5488\pm77$ K, $[\rm Fe/H]=0.34\pm0.08$ (\citealp{2015ApJ...808..187B}, \citealp{2018ApJ...852...46K}).

 \begin{table}
\begin{center}
\tablewidth{0pt}
\caption{Comparison between the parameters of theoretical models and spectroscopic parameters.\label{tbl-parameter_compare}}
\begin{tabular}{cccccc}
\tableline

  Parameters                        &Creevey et al.  &  Silva Aguirre et al. & Our results                           & Our results &   Spectroscopic parameters\\
                                    &  (2017)        &   (2017)     &   $Y_{\rm init} = 0.249+1.33Z_{\rm init}$&   $Y_{\rm init}$ as a parameter          & (Buchhave \& Latham 2015)  \\
\tableline
$M/(M_{\odot})$                    & $1.000\pm0.030$       &   $0.9861\pm0.0253$             &  $1.000_{-0.010}^{+0.005}$   & $1.02_{-0.03}^{+0.02}$ & \\
$R/R_{\odot}$                     & $0.930\pm0.009$       &    $0.9293\pm0.0083$             & $0.934_{-0.003}^{+0.001}$   & $0.940_{-0.009}^{+0.007}$  &\\
$L/L_{\odot}$                     & $0.64\pm0.03$        &    $0.7001\pm0.0473$              & $0.769_{-0.078}^{+0.039}$   &$0.656_{-0.097}^{+0.163}$  & \\
$T_{\rm eff}/K$                   &$5351\pm49$           &                                   & $5598_{-143}^{+64}$         & $5361_{-211}^{+320}$    &5488$\pm77$\\
$\log g$                          &$4.498\pm0.003$       &    $4.4953\pm0.0034$              &  $4.497_{-0.001}^{+0.001}$  &$4.500_{-0.004}^{+0.003}$  &\\
$t_{\rm age}(Gy)$                     &$4.57\pm0.36$         &    $5.1012\pm0.3253$              &$4.53_{-0.26}^{+0.30}$       &$4.90_{-0.63}^{+0.79}$    &\\
$\rm [Fe/H]$                      &$0.41\pm0.04$         &                                   & $0.270_{-0.032}^{+0.103}$   & $0.325_{-0.120}^{+0.062}$  &0.34$\pm0.08$\\
$R_{\rm CZ}/R$                    &                      &   $0.6927\pm0.0037$               &$0.685_{-0.005}^{+0.004}$    & $0.679_{-0.005}^{+0.012}$  &\\
$X_{C}$                           &                      &   $0.3979\pm0.0260$               &$0.382_{-0.014}^{+0.013}$    & $0.423_{-0.057}^{+0.041}$  &\\

\tableline
\end{tabular}
\end{center}
\end{table}

\begin{figure}
\gridline{\fig{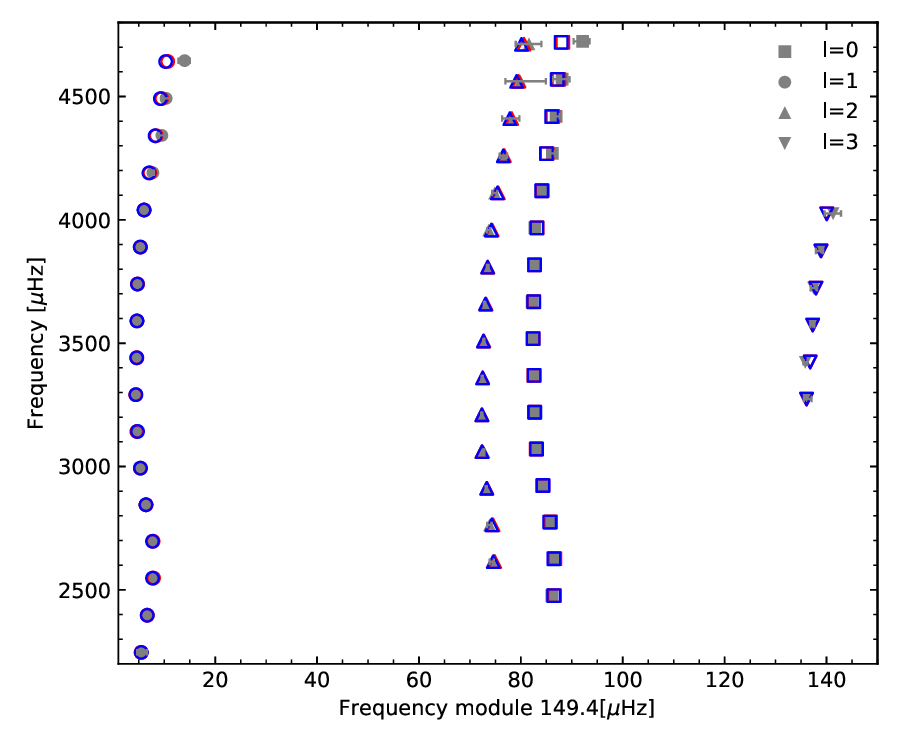}{0.5\textwidth}{}
          \fig{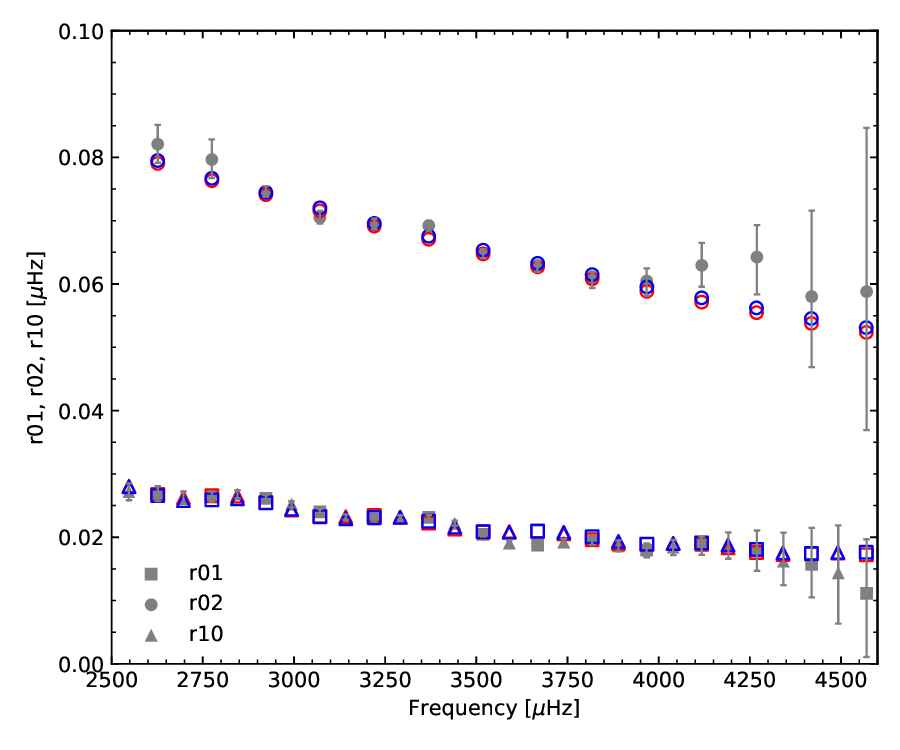}{0.5\textwidth}{}
           }

\caption{The $\rm \acute{E}$chelle diagram of observation data(filled gray points) of KIC 8006161 and the best-fit model with $Y_{\rm init} = 0.249+1.33Z_{\rm init}$(open red points) and $Y_{\rm init}$ as a free parameter (open blue points) and the corresponding ratio $r_{\rm ij}$ of small to large-frequency separations for $a=570$, and $b=130$. The errors of the observed frequencies and frequency ratios are from \cite{2017ApJ...850..110L}. The theoretical frequencies only consider the effects of small-scale magnetic fields, and do not include other physics associated with near-surface effects (i.e. Ball et al. 2016, Houdek et al. 2017, Schou \& Birch 2020). \label{fig:Echelle}}
\end{figure}

In order to show the matching details, we present the \'{E}chelle diagram of observation and the best-fit model with different $Y_{\rm init}$ and the corresponding ratio $r_{\rm ij}$ of small to large-frequency separations in Fig.\ref{fig:Echelle}. The filled gray points represent the observed data, while the open red and blue points show the results of best-fit model with $Y_{\rm init}= 0.249+1.33Z_{\rm init}$ and $Y_{\rm init}$ as a free parameter for $a=570$, and $b=130$, respectively. From Fig. \ref{fig:Echelle}, it can be seen that whether $Y_{\rm init}$ is set to $0.249+1.33Z_{\rm init}$ or treated as a free parameter, the $\rm\acute{E}$chelle diagram and corresponding frequency ratios of the best-fit models are essentially identical.
In the left panel, except for a few higher frequencies for $l = 0, l=1$, the most frequencies for the best-fit model with magnetic fields are in good agreement with the observation. In the right panel of Fig. \ref{fig:Echelle}, we present the distribution of ratios $r_{\rm ij}$ with frequencies. The errors of the frequency ratios in the observations are obtained from \cite{2017ApJ...850..110L}, and the theoretical ratios are calculated using Eq. (2) and (3). We find that for higher frequencies, the differences in the frequency ratios between the best-fit model with $Y_{\rm init}$ as $0.249+1.33Z_{\rm init}$ and as a free parameter and the observed data are larger. But due to the large errors in high-frequency observations, the majority of the theoretical ratio $r_{\rm ij}$ are within the observed error range.

 It is worth noting that these results do not take into account the other physics associated with near-surface effects. Comparing the observed oscillation frequencies with the theoretical ones, \cite{1988Natur.336..634C} found a systematical offset between them for the Sun. This offset was independent of the spherical degree $\ell$, but it increased with frequency (\citealp{1988Natur.336..634C}; \citealp{1988A&A...200..213D}). This is known as the near-surface effect of solar p-mode oscillation.

In general, the near-surface effect is mainly due to the interactions between p-mode oscillations and turbulent convection near the top of the stellar convection zone. \cite{1999A&A...351..689R} considered the averages of hydrodynamical simulations and found that turbulent pressure provided additional support against gravity. This would result in higher turning points of high-frequency modes. Subsequently, \cite{2016A&A...592A.159B} used averaged three-dimensional radiation hydrodynamics simulation (\citealp{2013A&A...558A..48B}) to replace the near-surface structure of dwarfs with different spectral types. Houdek et al. considered the averaged structure of a 3D hydrodynamical simulation and the physical effects of non-adiabaticity and turbulent pressure (\citealp{2017MNRAS.464L.124H}; \citealp{2020A&A...638A..51S}; \citealp{2021MNRAS.500.4277J}). They got the frequency deviations with $\leq 3 \rm\mu Hz$ for the sun. In this work, we only consider the influence of small-scale magnetic fields on the oscillation frequencies of solar-like star KIC 8006161 and do not include the effects of non-adiabaticity and turbulent pressure.

\subsection{Small-scale magnetic fields}
\begin{figure}[ht!]
\plotone{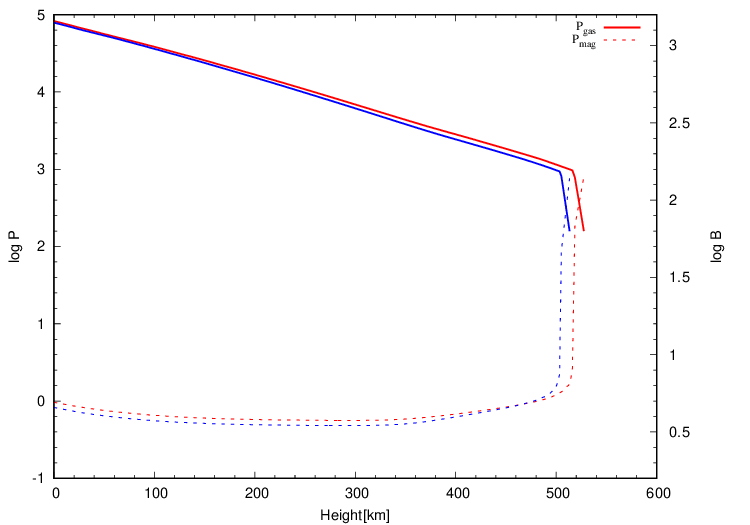}
\caption{The distributions of the gas pressure and the magnetic pressure, as well as the fields strength for the best-fit model with $Y_{\rm init} = 0.249+1.33Z_{\rm init}$(red lines) and $Y_{\rm init}$ as a free parameter (blue lines) by adopting the grey atmosphere. \label{fig:logP_height}}
\end{figure}

The magnetic fields and the acoustic waves are coupled with each other. Fig. \ref{fig:logP_height} presents the distributions of gas pressure and magnetic pressure with height for the best-fit model with $Y_{\rm init} = 0.249+1.33Z_{\rm init}$(red lines) and $Y_{\rm init}$ as a free parameter (blue lines). When the stellar magnetic pressure increases with height, the gas pressure decreases. This subsequently leads to a decrease in density and an increase in acoustic speed. According to \cite{2002ApJ...564..508R} and \cite{2007AN....328..286C}, the acoustic waves are reflected and transmitted at a position where the gas pressure equals magnetic pressure. We use this condition to calculate the magnetic field strength and determine the location of the magnetic-arch splicing layer for the best-fit model with $Y_{\rm init} = 0.249+1.33Z_{\rm init}$ and $Y_{\rm init}$ as a free parameter. We find that the magnetic field strength for the best-fit model with $Y_{\rm init} = 0.249+1.33Z_{\rm init}$ is $96$ G, and the height of the magnetic-arch splicing layer is $522$ km. For the best-fit model with $Y_{\rm init}$ as a free parameter, the magnetic field strength is 89 G, and the height of the splicing layer is about 510 km. This result also indicates that whether $Y_{\rm init}$ is varied with $\triangle Y/\triangle Z$ or treated as a free parameter, the best-fit model has a similar stellar small-scale magnetic field strength. In addition, there is a large area below the magnetic-arch splicing layer in Fig. \ref{fig:logP_height} with a very low magnetic field strength. This is very similar to the configuration of magnetic voids just below the arches observed in the magneto-hydrodynamic (MHD) simulations (\citealp{2005ESASP.596E..65S,2006ASPC..354..345S}) .

 KIC 8006161 is a main sequence star with a higher metallicity. Table \ref{tbl-parameter_compare} presents the stellar parameters for KIC 8006161. We can get that KIC 8006161 is a star with $\sim1M_{\odot}$ and a smaller stellar radius. The stellar age is similar to that of the sun. Although the origin mechanism of small-scale magnetic fields is still controversial, stellar convection plays an important role in their generation. We compare the distribution of stellar physical properties in the convection zone between KIC 8006161 and the sun.

\begin{figure}
\gridline{\fig{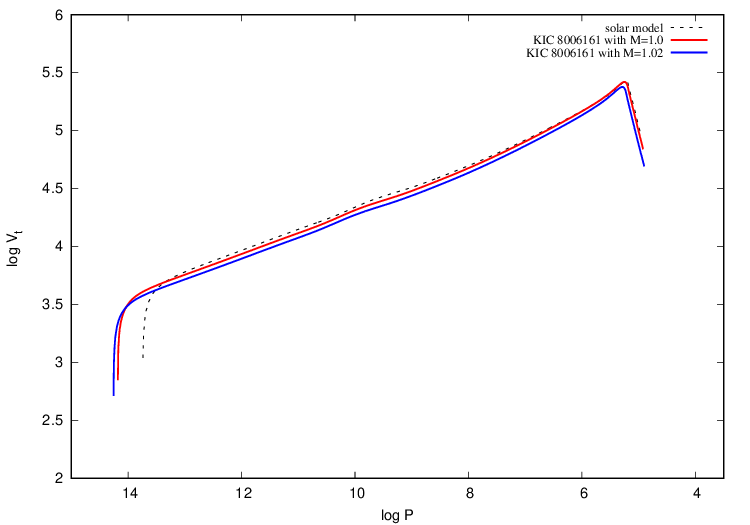}{0.5\textwidth}{}
          \fig{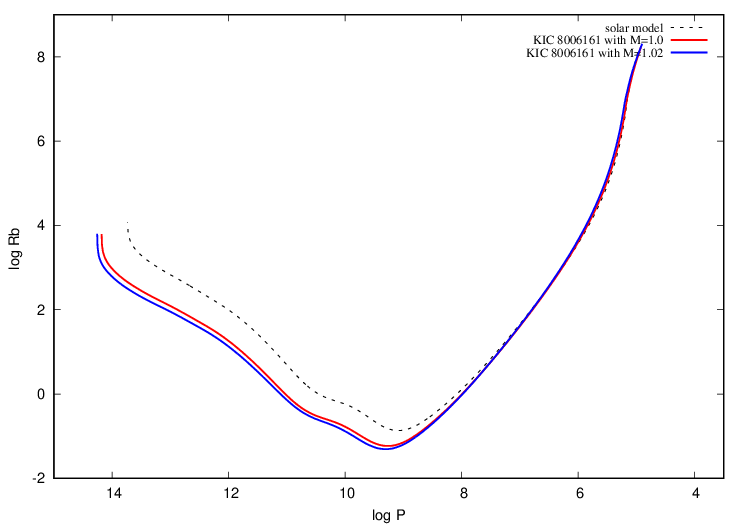}{0.5\textwidth}{}
           }

\caption{The distribution of physical quantities in the convection zone between KIC 8006161 (solid lines) and the solar model (dashed lines) from \cite{2021ApJ...916..107L}. The red lines represent the results of the best-fit model with $Y_{\rm init}=0.249+1.33Z_{\rm init}$, and the blue lines represent the results of the best-fit model with $Y_{\rm init}$ treated as a free parameter. The left panel is the distribution of typical velocity of turbulent motions with $\log P$ in the convection zone. The right panel is the distribution of typical size of convective rolling cells. \label{fig:comparisons_two stars}}
\end{figure}

   We use the results of the solar model from \cite{2021ApJ...916..107L}, and present a comparison of the typical velocity of turbulent motions between two stars in the left panel of Fig. 6. The black dotted line represents the result of the solar model, while the red and blue lines represent the results of a best-fit model with $Y_{\rm init}=0.249+1.33Z_{\rm init}$ and $Y_{\rm init}$ as a free parameter, respectively. It can be seen that KIC 8006161 has a deeper convection zone. The best-fit model with $Y_{\rm init}=0.249+1.33Z_{\rm init}$ and the solar model have nearly identical maximum  turbulence velocities (about $2.6\times10^{5}$ cm s$^{-1}$), while the best-fit model with $Y_{\rm init}$ as a free parameter exhibits slightly lower maximum turbulence velocities (about $2.4\times10^{5}$ cm s$^{-1}$). Except for the different location of convection zone bottom, the distribution of typical velocity of turbulent motions is basically the same for the solar model and the best-fit model with $Y_{\rm init}=0.249+1.33Z_{\rm init}$. We also compare the typical size of convective rolling cells in the right panel of Fig. \ref{fig:comparisons_two stars}, which is expressed by Eq. (58) from \cite{2012ApJ...756...37L}. We can get that the typical size of convective rolling cells $R_{\rm b}$ in a $k-w$ model (\citealp{2012ApJ...756...37L}) is nearly consistent with each other just below the stellar photosphere. On the stellar surface, convective rolling cells can reach scales of more than one thousand kilometers.

 In summary, although there are some differences in the fundamental parameters between KIC 8006161 and the solar model, the characteristics of the convection zone (i.e. maximum velocity of turbulence and typical size of convective rolling cells) are very similar to each other. Our theoretical model suggests that the small-scale magnetic field strength of KIC 8006161 is very similar to that of the sun.

\section{Summary}
 For solar-like stars, there is a convective envelope on the stellar surface. The small-scale magnetic fields may be widely distributed on the surface of solar-like stars. Because small-scale magnetic fields are ubiquitous on stellar surface (\citealp{2021ApJ...916..107L}; \citealp{2004Natur.430..326T}), they store a amount of magnetic energy. This would affect the heating of the corona and the acceleration of the stellar wind (\citealp{1994ASSL..189.....S}; \citealp{1998Natur.394..152S}; \citealp{2003ApJ...597L.165S}). In addition, the small-scale magnetic fields are coupled with the acoustic wave, influencing the propagation speed of the wave (\citealp{2002ApJ...564..508R}; \citealp{2003ApJ...599..626B};\citealp{2007AN....328..286C}). This also provides us with an opportunity to use asteroseismology to constrain the small-scale magnetic field of solar-like star.

 In this work, considering the effects of small-scale magnetic fields in the photosphere on the oscillation frequencies, we attempt to use the observed oscillation frequencies to constrain the stellar fundamental parameters for KIC 8006161, and especially get the strength of the small-scale magnetic fields. In our model, we assume that the surface effects are caused by the small-scale magnetic fields of the star. We approximate the influence of the magnetic field by incorporating magnetic pressure into the grey-atmosphere model. Due to the impact of turbulent pressure, which provides additional support against gravity(\citealp{1999A&A...351..689R}; \citealp{2017MNRAS.464L.124H}; \citealp{2020A&A...638A..51S}; \citealp{2021MNRAS.500.4277J}), the magnetic field strength obtained from our models may approach the upper limit of KIC 8006161. Our main conclusions are as follows:

1. Considering the effects of small-scale magnetic fields in the photosphere, we derived the stellar fundamental parameters with a $\chi^{2}$-minimum method of iterative computation for KIC 8006161, which are listed in Table \ref{tbl-parameter_compare}. We use two methods to determine the initial helium abundance $Y_{\rm init}$: one uses the chemical enrichment law $\triangle Y/\triangle Z = 1.33$, and the other treats $Y_{\rm init}$ as a free parameter. We find that considering the effects of small-scale magnetic fields in the photosphere, the stellar fundamental parameters obtained by two methods are consistent with the results from spectroscopic observations. In addition, the initial helium abundance $Y_{\rm init}$ for the best-fit model with $Y_{\rm init}$ as a free parameter is lower than the value obtained from the relation $Y_{\rm init}=0.249+1.33Z_{\rm init}$. The corresponding stellar mass is also larger, $\sim 1.02M_{\odot}$. The other fundamental parameters (i.e. radius R, $\log g$, $R_{\rm CZ}/R$, $X_{\rm C}$) are completely consistent for both models.

2. Considering the effects of small-scale magnetic fields regardless of the value of $Y_{\rm init}$, most of the theoretical frequencies are in agreement with the observational ones. Meanwhile, most of the ratio $r_{\rm ij}$ of small to large-frequency separations for the best-fit model with $Y_{\rm init} = 0.249+1.33Z_{\rm init}$ and $Y_{\rm init}$ as a free parameter also agree with the observational ones. This also implies that our best-fit model is in good agreement with observations from the star's interior to its surface.

3. The small-scale magnetic field strength for KIC 8006161, independently derived from our model with $Y_{\rm init} = 0.249+1.33Z_{\rm init}$ and $Y_{\rm init}$ as a free parameter, is approximately $96$ G and 89 G, respectively. The corresponding locations of the magnetic-arch splicing layer are about $522$ km and 510 km, respectively. Since the stellar convective properties for KIC 8006161 are similar to those of the solar model from \cite{2021ApJ...916..107L}, regardless of the value of $Y_{\rm init}$, the magnetic field strength is also similar to that of the sun.


\begin{acknowledgments}

This research is supported by National Natural Science Foundation of China (grant No. 12288102 and 12133011), National Key R\&D Program of China (grant No. 2021YFA1600400/2021YFA1600402), the B-type Strategic Priority Program of Chinese Academy of Sciences (grant No. XDB41000000), National Natural Science Foundation of China (grant No. 12273104), Yunnan Fundamental Research Projects (grant No. 202401AS070045), International Center of Supernovae at the Yunnan Key Laboratory (grant No. 202302AN360001/202302AN36000101) and the West Light Foundation of the Chinese Academy of Sciences. T. W. thanks for financial support from Youth Innovation Promotion Association of Chinese Academy of Sciences, and Yunnan Ten Thousand Talents Plan Young \& Elite Talents Project.
\end{acknowledgments}

\bibliography{sample631}{}
\bibliographystyle{aasjournal}



\end{document}